# Anomalous Nernst effect in Co thin films under laser irradiation

Soichiro Mochizuki,[1] Itaru Sugiura,[2] Tetsuya Narushima,[1] Teruo Ono,[2,3]
Takuya Satoh,[1] and Kihiro T. Yamada[1,*]

[1]*Department of Physics, Institute of Science Tokyo, Tokyo 152-8551, Japan*

[2]*Institute for Chemical Research, Kyoto University, Uji, Kyoto 611-0011, Japan*

[3]*Center for Spintronics Research Network, Institute for Chemical Research, Kyoto University, Uji, Kyoto, 611-0011, Japan*

[*]Corresponding author: yamada@phys.sci.isct.ac.jp

**Abstract**

The anomalous Nernst effect (ANE) generates electromotive forces transverse to temperature gradients and has attracted much attention for potential applications into alternative thermoelectric power generators. ANE efficiency is generally characterized by uniform temperature gradients in a steady state prepared by heaters. However, although focusing laser beams on a magnetic film can form much larger temperature gradients, the laser irradiation method has not been sufficiently considered for quantifying the ANE coefficient due to the difficulty in estimating the localized in-homogeneous temperature gradients. In this study, we present a quantitative study of ANE in Ru(5 nm)/Co($t_{Co}$) ($t_{Co}$ = 3, 5, 7, 10, 20, 40, and 60 nm) bilayers on sapphire (0001) substrates by combining a laser-irradiation approach with finite-element analysis of temperature gradients under laser excitation. We find that the estimated ANE coefficients are consistent with previously reported values and one independently characterized using a heater. Our results also reveal the advantages of the laser-irradiation method over the conventional method using heaters. Intensity-modulated laser beams can create ac temperature gradients as large as approximately $10^3$ K/mm at a frequency of tens of kilohertz in a micrometer-scale region.

## I. INTRODUCTION

Spin caloritronics [1,2], which studies the spin transport properties driven by heat currents in magnetic materials, has developed rapidly since the discovery of the spin Seebeck effect [3]. The anomalous Nernst effect (ANE) [4] generates electromotive forces ($E_{ANE}$) in a cross-product direction of magnetization ($\mu$) and temperature gradient ($\nabla T$) as follows:

$$E_{ANE} = Q_{ANE} (\mu \times \nabla T), \quad (1)$$

where $Q_{ANE}$ is the ANE coefficient. While Seebeck generators produce electromotive forces in the same direction as $\nabla T$, the $E_{ANE}$ is generated orthogonal to the $\nabla T$. This transverse configuration can be used to fabricate thin flexible thermoelectric power generators. However, their thermoelectric conversion efficiencies are insufficiently high for practical power generators. To increase conversion efficiency, material developments recently have focused on magnetic alloys, magnetic hybrid structures, and topological magnets [5–17].

Uniform temperature gradients prepared using heaters are generally used to study ANE and the spin Seebeck effect [18–21]. By contrast, temperature gradients can be formed using laser absorption [22–24]. The temperature gradients formed using laser irradiation are localized, which enables visualization of the local magnetization direction by measuring the generated electric voltages [25–31]. Moreover, because visible laser beams exponentially decay by tens of nanometers inside a metallic film, the temperature gradient should be enormous compared to when a temperature difference is prepared using heaters across a sub-millimeter-thick substrate. However, the difficulty in estimating temperature gradients under laser excitation has led to considerable variation in the simulated temperature gradients, which thereby impeding the widespread use of the laser-irradiation method in quantitative studies of ANE. In existing works [22–27], only heat conduction has been simulated using finite-element methods. A more accurate approach requires simulating light propagation especially for multilayered thin films on substrates where light reflections occur at interfaces. More detailed experimental and simulation studies are essential to further promote the use of the laser irradiation method to study spin caloritronics.

In this study, we present a quantitative study of the ANE in Co thin films using laser irradiation. Because the laser-absorption profile is sensitive to film thicknesses of tens of nanometers, we exploited the Co thickness of $Q_{ANE}$ to ascertain the applicability of the laser-irradiation method. To estimate the values, the transverse electric voltages were measured when the magnetic film was excited by intensity-modulated laser beams. Moreover, finite-element methods were employed to simulate not only the spatial profiles of light electric-field profile but also those of temperature inside the Co film. We find a quantitative and qualitative agreement of the $Q_{ANE}$ values with previously reported values and our value characterized using a heater. Furthermore, our results reveal the multiple merits of the laser-irradiation method compared to the conventional method of using heaters.

## II. METHODS

### A. Experimental configuration

We deposited bilayers consisting of Ru(5 nm)/Co($t_{Co}$) ($t_{Co}$ = 3, 5, 7, 10, 20, 40, and 60 nm) on *c*-plane sapphire substrates by dc sputtering. The Ru capping layer was selected owing to its tolerance to oxidization. By standard photolithography techniques, the bilayers were processed to a Hall cross of (*w*, *l*) = (0.40 mm, 0.40 mm), where *w* and *l* denote the width and length of the Hall cross, respectively. Electrical contacts consisting of Au(20 nm)/Cu(80 nm)/Ti(5 nm) by rf sputtering. The bottom Ti layer was used to improve adhesions of electrodes to sapphire substrates. The electrical contacts were connected with gold wires using a wire bonder.

Figure 1 shows a schematic of the experimental setup. A diode-pumped solid-state green laser with a wavelength of $\lambda$ = 532 nm was used for the small beam divergence angle. The laser intensity was modulated stably using the optics shown in Fig. 1. Linearly polarized laser beam with a polarization angle of +45° from a Glan Taylor prism propagates into a photo-elastic modulator (PEM) which was set to give a retardation of $\gamma = \gamma_0 \sin\Omega t$ to the *X*-polarized component. The polarization-modulated laser beam passed through a half-wave plate, of which the fast axis was rotated by +22.5° from the *X*-axis. Then, the *Y*-polarized component was then extracted using another Glan Taylor prism. In this current configuration, the polarization state of the laser output is described using the Jones vector, as follows [32]:

$$\vec{E} = \begin{pmatrix} 0 & 0 \\ 0 & 1 \end{pmatrix} \frac{1}{\sqrt{2}} \begin{pmatrix} 1 & 1 \\ 1 & -1 \end{pmatrix} \begin{pmatrix} e^{i\gamma} & 0 \\ 0 & 1 \end{pmatrix} \frac{1}{\sqrt{2}} \begin{pmatrix} 1 \\ 1 \end{pmatrix} = \frac{1}{2} \begin{pmatrix} 0 \\ e^{i\gamma} - 1 \end{pmatrix}. \quad (2)$$

Using Bessel functions of the first kind for $n^{th}$ integer orders, $J_n(\gamma_0)$, the intensity of the output, $I_{out}$, is expressed as follows:

$$I_{out} = \vec{E}^* \cdot \vec{E} = \frac{1}{2} - \frac{1}{2}\cos(\gamma_0 \sin\Omega t)$$

$$= \frac{1}{2} - \frac{1}{2} \cdot J_0(\gamma_0) + 2J_0(\gamma_0)\cos2\Omega t + J_1(\gamma_0)\cos4\Omega t + \cdots \quad (3)$$

The retardation of the PEM was set at $\gamma_0 = \pi$. In this case, $I_{out}$ becomes

$$I_{out} = 0.6521 - 0.4854\cos2\Omega t - 0.1514\cos4\Omega t + \cdots. \quad (4)$$

The intensity ratio of the dc, 2Ω, and 4Ω components is 1:0.744:0.232. For calibrating the PEM, we measured the dc and 2Ω voltages from a photodetector using a multimeter and a lock-in amplifier, respectively. The driving frequency of the PEM was fixed at $\Omega/2\pi$ = 42 kHz. The intensity modulation method using a PEM is more accurate and stable than a mechanical method using an optical chopper. Using an objective lens with a magnification of ×2, the laser output was focused on the center of the Hall cross to maintain symmetric in-plane temperature gradients in space. We directly checked using a camera that the focused beam had a Gaussian intensity profile with a $1/e^2$ radius, $\sigma$ = 25 μm. The magnetization of Co was saturated in a direction titled by $\varphi$ from the *X*-axis within the film plane using a magnetic field generated by an electromagnet mounted on a rotation stage. During the illumination of the laser beam, we measured $2\Omega$ voltages between the counter electrodes using a lock-in amplifier.

### B. Simulation condition

COMSOL Multiphysics 6.2 [33] was used to evaluate the temperature gradients inside the Co films created by laser irradiation. The spatial profile of the light electric field was simulated using the Electromagnetic Waves, Beam Envelopes interfaces of the Wave Optics Module [34]. Moreover, the spatial temperature profile was simulated using the Heat Transfer in Solids interface of the Heat Transfer Module [35]. Figure 2 shows the simulation geometry composed of the sapphire substrate layer, Ru and Co layers, and air 1 and air 2 layers. Note that the definition of the coordinate system is changed in the following calculations: 9 ∥ ≼? ∥ −@A ∥ −⓪. For the constraints of the Wave Optics Module, we simulated the spatial profile of the light electric field in the substrate, Ru/Co, and air 1 layers. The spacious air 2 layer and the substrate layer act as heat baths when calculating the spatial temperature profile. Using the geometrical condition of Fig. 2(a) and refractive indexes listed in Table 1, we first simulated the spatial profiles of light electric field when Gaussian laser beam with a $1/e^2$ radius 8 = 25 μm and $\lambda$ = 532 nm propagates from the air to metallic layers. Here, we used typical bulk values for the refractive indexes at $\lambda$ = 532 nm. Subsequently, the spatial temperature profile was simulated in the axisymmetric geometry of Fig. 2(b) using the thermal conductivities of the Ru and Co films listed in Table 1. The thermal conductivities were estimated on the basis of their resistivities, as discussed later. The temperatures at the bottom of the substrate layer and the top of the air 2 layer are fixed at room temperature (293.15 K). The sample layers and the substrate layer are made wider than the air layer to allow the heat flow in the in-plane direction. We verified that the convergence and reliability did not change by further expanding the thicknesses of the air 2 and substrate layers. While the temperature profile of the substrate layer changes with a further increase in thickness, the temperature profile of the thin Co layer is hardly affected. Simulating the time evolution of the temperature profiles in three-dimensional space requires substantial computational cost and time. Therefore, we calculated the time evolutions of temperature profiles in the axisymmetric geometry using energy-dissipation profiles calculated in the two-dimensional system.

### III. RESULTS AND DISCUSSION

#### A. Magnetization, resistivity, and thermal conductivity

We characterized the $t_{Co}$ dependence of the saturation magnetization, $\mu_0 M_s$, shown in Fig.3(a) using a superconducting quantum interference device and a vibrating-sample magnetometer. With increasing $\mu_0 M_s$ approaches the bulk value of 1.8 T [36]. The sheet resistances, $R_s$, of the fabricated devices, shown in Fig. 3(b) were measured using four-terminal sensing. The nonlinear $t_{Co}$ dependence of $R_s$ was used to estimate the resistivities of Ru ($\rho_{Ru}$) and Co ($\rho_{Co}$) layers. We assumed that the bilayer was equivalent to a parallel circuit consisting of the resistances of the Ru and Co layers, and the resistivities were constant in the Co and Ru thin films. Because the width and length of the Hall devices are equal, the sheet resistance is given by

$$R_s = \frac{\rho_{Ru}\rho_{Co}}{\rho_{Ru}t_{Co} + \rho_{Co}t_{Ru}}, \qquad (5)$$

where the $t_{Ru}$ is the Ru thickness of 5 nm. Fitting the $t_{Co}$ dependence of the $R_s$ shown in Fig. 3(b) with Eq. (5) gives $\rho_{Ru} = (5.15 \pm 0.49) \times 10^{-7}$ $\Omega$ m and $\rho_{Co} = (4.06 \pm 0.32) \times 10^{-7}$ $\Omega$ m. Furthermore, by applying the Wiedemann-Franz law [37] to the resistivities, we estimated the thermal conductivities of the Ru and Co layers to be $14.3 \pm 1.4$ W m$^{-1}$K$^{-1}$ and $18.1 \pm 1.4$ W m$^{-1}$K$^{-1}$, respectively. The calculated thermal conductivities are comparable to previously-reported values of thin metals, including Co [38]. The estimated thermal conductivities were used in the simulations.

**B. The anomalous Nernst effect induced by laser irradiation**

Figure 4(a) shows the magnetic hysteresis of the electric voltages ($V$) as a function of the magnetic field ($H$) at $\varphi = 90°$ and an average laser power ($P$) of 70 mW for $t_{Co} = 10$ nm. This clear hysteresis indicates that the electromotive force reflects the magnetization of the Co layer. Furthermore, we define the voltage jump by magnetization reversals as $\Delta V = [V(H^+) - V(H^-)]/2$ at $\mu_0 H^{\pm} = \pm 80$ mT. The measured $\varphi$ dependence of $\Delta V$ for $t_{Co} = 10$ nm is shown in Fig. 4(b). For all the bilayers, the $\varphi$ dependences of $\Delta V$ can be very well fitted with a sine function $\Delta V_{fit}\sin\varphi$. The values of $\Delta V_{fit}$ at $P = 70$ mW are plotted versus $t_{Co}$ in Fig. 4(b). We observe a peak in the $\Delta V_{fit}$ at $t_{Co} = 7$ nm. Note that the $\Delta V$ linearly depends on $P$ within the employed range of $P$ at all the conditions of $t_{Co}$. For example, refer the case where $t_{Co} = 7$ nm [Fig. 4(d)].

**C. Simulated spatial profiles of light electric fields and temperature**

We simulated the spatial profiles of light electric field and temperature when the laser beam with $\lambda = 532$ nm propagates into the bilayer on the sapphire substrate. Figure 5(a) shows the spatial profiles of the light electric-field amplitude, |**E**|, at $t_{Co} = 60$ nm in the steady state. The laser beam was incident from air to the surface of the Ru layer. The light electric field is partially reflected at the interface between the air and Ru layers. Although the interference between the reflected and incident light generates standing light waves in the air layer, it does not affect the amplitude of the transmitted light electric field. The transmitted-light electric field decays exponentially inside the metallic layers [Fig. 5(b)]. Then, we obtained the spatial profile of the energy dissipation density shown in the bottom of Fig. 5(b). The discontinuity of the energy dissipation density at the interface is ascribed to the difference in the extinction coefficients. Note that the Wave Optics Module cannot handle pulsed and amplitude-modulated light electric fields. Instead, in the Heat Transfer Module, we introduced heat sources, of which the caloric generation oscillates at $2\Omega/2\pi$ while keeping the spatial profile of energy dissipation density (absorbed laser intensity). Because the following simulations considered only the laser power oscillating at $2\Omega/2\pi = 84$ kHz, the amplitude of the $2\Omega$ component ($P_{2\Omega}$) of laser power is obtained by multiplying the $P$ by a factor of 0.744.

Figure 5(c) shows the spatial temperature profile at $P_{2\Omega} = 52.1$ mW when the temperature difference between the top and bottom surfaces of the Co layer ($\Delta T$) is maximized. Figure 5(d) presents a magnified view of Fig. 5(c), in which we defined a cut-line (broken line) along the center. The temperatures along the cut-line change nonlinearly owing to the exponential decay of the light intensity [Fig. 5(e)]. Moreover, we evaluated the nonlinear temperature gradient by integrating the temperature gradients, $R = \int_{V(W_M \times W_{Co})}^{V W_M} \nabla T(A) $, which are divided by $t_{Co}$. We note that the feasible temperature gradients can reach approximately $10^3$ K mm$^{-1}$, which is two orders of magnitude larger than those when using heaters (~10 K mm$^{-1}$) [7,9,14,18,21]. The $t_{Co}$ dependence of $R/Z_{H''}$ in Fig. 5(f) has a local minimum at $t_{Co} = 20$ nm. When the Co thickness is thin, the back reflected light waves from the Co/sapphire interface tend to cancel out the temperature gradients inside the Co layer. With increasing $t_{Co}$, the incident light waves tend to decay only within the Co layer before reaching the interface. Indeed, the position of the local minimum varies depending on the extinction coefficient of Co (see APPENDIX A). Therefore, the local minimum determined mainly by the back reflection from the Co/sapphire interface and light decay within the Co layer.

For comparison, we also simulated the spatial temperature profile in the two-dimensional geometry shown in Fig. 2(a) (see APPENDIX B). The calculated value of $R/Z_{H''}$ in the two-dimensional geometry is plotted as a function of $t_{Co}$ in Fig. 9(d). The difference in $R/Z_{H''}$ between the axisymmetric and two-dimensional cases is only a few percent. We also observe an agreement between the horizontal temperature and light intensity profiles [Fig. 9(e)]. These results indicate that the heat transport occurs predominantly in the thickness direction. The laser intensity contains dc and multiple even-frequency components whereas the simulation considered only the $2\Omega$ components. Nevertheless, when dc, $2\Omega$, and $4\Omega$ components of laser intensity are considered in the simulation, we find no significant emergences of other frequency components by the distortion of the $\Delta T$ waveforms (see APPENDIX C). The laser-irradiation method can steadily and stably modulate the temperature gradient at a frequency of tens of kilohertz, which is applicable to phase-detection techniques for accurately characterizing the magneto-thermoelectric effects.

### D. Geometrical correction of the anomalous Nernst effect induced by focused laser beams

Because electromotive forces are generated only in the region irradiated by laser beams, we must correct the generated electric voltages shown in Fig. 4(c) by considering the measurement geometry to calculate the values of $Q_{ANE}$. The geometrically corrected electric voltage, $'\bar{A}_]$ , is given by

$$S_{xy}' = -\frac{\delta^0}{2\pi Z_H''} a\mu b_{ANE} c \frac{VW_M}{V(W_M \times W_0)} T_A \nabla T(A). \tag{6}$$

APPENDIX D provides detailed derivations of Eq. (6). To confirm the validity of Eq. (6), where $S_{xy}'$ is inversely proportional to $w$, we fabricated Hall devices with $w = l = 0.10, 0.20, 0.30$ and $0.40$ mm using the Ru(5 nm)/Co(60 nm) bilayer. Figure 6(a) shows $V$ as a function of $H$ at $\varphi = 90°$ and $P = 70$ mW for $w = 0.10, 0.20, 0.30$, and $0.40$ mm. Moreover, the linear dependence of $\Delta V$ on $1/w$ shown in Fig 6(b) indicates that Eq. (6) is valid. Notably, Eq. (6) indicates that the electric voltage generated by laser irradiation is optimized when the device size $l$ becomes as large as the spot size $\delta$. Hence, the laser-irradiation method can generate detectable electric voltages in micrometer-sized devices by focusing the laser beams, whereas heater methods require a millimeter-sized device owing to the small temperature gradient. The laser-irradiation method can expand the range of thermoelectric materials that can be investigated, including micrometer-sized single crystals and cleaved van der Waals layered materials [39–41].

### E. Calculation of the coefficient of the anomalous Nernst effect $Q_{ANE}$

Figure 7 shows the $t_{Co}$ dependence of $Q_{ANE}$, which was calculated by substituting the characterized values of $\mu_0 M_s$ [Fig. 3(a)], $S_{xy}'$ ($= \Delta V_{fit}$) [Fig. 4(c)], and $\nabla T$ [Fig. 5(f)] to Eq. (6). The inset of Fig. 7 shows the transverse Seebeck coefficient, $S_{xy}$. The error bars represent the uncertainty of the thermal conductivity of Ru and Co. The estimation of $Q_{ANE}$ shown in Fig. 7 ignores the spin Seebeck effect for the small spin Hall angle of Ru [42]. $Q_{ANE}$ increases with a reduction in $t_{Co}$ and is maximized at $t_{Co} = 7$ nm. Reference [21] reported a similar dependence of $Q_{ANE}$ on the thickness of the ferromagnetic metal. This increasing tendency can be explained by the increasing contribution of impurity scattering or the intrinsic ANE due to the modification of the electronic band structure, based on the Mott relationship [43,44]. In contrast, the reduction in $Q_{ANE}$ at less than 5 nm may be ascribed to the decrease in the Curie temperature. The values of $Q_{ANE}$ shown in Fig. 7 have the same order of magnitude as those reported in previous studies ranging from 0.02 to 0.40 µV K$^{-1}$ T$^{-1}$ [21,45,46]. Furthermore, we measured the ANE at $t_{Co} = 7$ nm using a conventional method with a heater for comparison (see APPENDIX F). The estimated $Q_{ANE}$ is 0.22 µV K$^{-1}$ T$^{-1}$ compared with 0.14 µV K$^{-1}$ T$^{-1}$ in Fig. 7. The consistency between the laser-irradiation and heater experiments also indicates that the laser irradiation method is usable in the quantitative study of ANE when combined with finite element simulation. We observe a systematic error in $Q_{ANE}$ of approximately 40%, which is attributed mainly to the selection of material parameters in the simulation. For example, at $t_{Co} = 10$ nm, the variation of the thermal conductivity of Co by +7.9% ($-7.9\%$) resulted in a change of $\Theta/t_{Co}$ by $-6.9\%$ (+8.4%), respectively. The values of $\Theta/t_{Co}$ also depend on the extinction coefficient of Co (see APPENDIX A). We additionally note that $\Theta/t_{Co}$ should change by the Fresnel coefficients of the Ru/air and Ru/Co interfaces. The consideration of actual material parameters in the simulations should lead to a more accurate estimation of $Q_{\mathrm{ANE}}$.

## IV. CONCLUSION

We quantitatively studied the thickness dependence of the anomalous Nernst effect induced by laser irradiation in Ru(5 nm)/Co($t_{\mathrm{Co}}$) ($t_{\mathrm{Co}}$ = 3, 5, 7, 10, 20, 40, and 60 nm) bilayers. The measured transverse electric voltages and simulated temperature gradients were used to calculate the anomalous Nernst coefficients. The thickness dependence of the anomalous Nernst coefficients quantitatively and qualitatively reproduces those in previous reports and one independently characterized using homogeneous temperature gradients prepared by a heater. This study also highlights that temperature gradients as large as approximately $10^3$ K/mm can be created by focusing laser beams on a micrometer-sized area while stably modulating the intensity at a frequency of tens of kilohertz. These advantages of the laser-irradiation method are beneficial for understanding the physics of the anomalous Nernst effect and exploring novel thermoelectric materials.

**ACKNOWLEDGMENTS**

We thank Takafumi Obayashi, Tomoya Itoh, and Tetsuma Mandokoro for their technical assistance and Hiro Munekata for his fruitful discussion. This study was partially supported by JSPS KAKENHI (Grant Nos. JP22K14588, JP25H00612, JP24K00938, JP23K22425, JP21H01032, JP20H05665, JP24H00007, and JP23H01984), the Frontier Photonic Sciences Project (Grant Nos. JP01212307, and JP01212405), MEXT X-NICS (Grant No. JPJ011438), ASUNARO Grant, JST CREST (Grant No. JPMJCR24R5), and the Collaborative Research Program of the Institute for Chemical Research, Kyoto University.

**APPENDIX A: ADDITIONAL SIMULATIONS WHEN VARIATING EXTINCTION COEFFCIENTS**

To seek the origins of the nontrivial $t_{Co}$ dependence of $\Theta/t_{Co}$ in Fig. 5(f), we calculated $\Theta/t_{Co}$ while varying the extinction constant (imaginary part of the refractive index) of Co, $k_{Co}$, at each $t_{Co}$, as shown in Fig. 8. We first found that the local minimum in Fig. 5(f) disappears at $k_{Co} = 0$, indicating that the nontrivial behavior originates from the laser absorption by Co. In addition, the values of $\Theta/t_{Co}$ and the positions of the local minima vary depending on $k_{Co}$. When the Co film is thin ($t_{Co}$ = 3 nm–20 nm), the electric-field amplitudes in the Ru layer depend on $t_{Co}$ and gradually converge as $t_{Co}$ increasing further. The variation of the electric-field amplitude in the Ru layer is attributed to reflected light waves returning to the Ru from the Co/sapphire interface. Hence, the dissipation energy density in the Ru layer depends on $t_{Co}$ for thin Co films and gradually converges with increasing $t_{Co}$. Of note, the reflected light also tends to cancel out the temperature gradients in the Co layer. Moreover, $\Theta/t_{Co}$ begins to increase at a specific value of $t_{Co}$ because the electric-field amplitude at the Co/sapphire interface decays with increasing $t_{Co}$. Therefore, the nontrivial $t_{Co}$ dependence in Fig. 5(f) results likely from the combined effects of optical propagation and laser absorption.

**APPENDIX B: COMPARISION BETWEEN TWO-DIMESIONAL AND AXISYMMETRIC CASES**

We also simulated the temperature profile in the two-dimensional geometry [Fig. 2(a)] as shown in Fig. 9(a). Figures 9(b) and 9(c) present the magnified view and temperature difference along the cut line, respectively. Figure 9(d) compares the $t_{Co}$ dependence of $\Theta/t_{Co}$ in the two-dimensional geometry to that in the axisymmetric geometry [Fig. 5(f)]. The deviation of $\Theta/t_{Co}$ between the two cases is approximately 2%. Moreover, we plot the in-plane temperature profile at $z$ = 35 nm for $t_{Co}$ = 60 nm, calculated in the two-dimensional system, together with the normalized square of light electric field, $|\mathbf{E}|^2$, at the sample surface in Fig. 9(e). Fitting the in-plane temperature profile with a Gaussian yielded $\sigma$ = 25.5 μm, which is broaden by only 2% in comparison with the light intensity (8 = 25 μm). These results indicate that heat flows primarily along the thickness direction rather than in the in-plane direction.

**APPENDIX C: TEMPERATURE MODULATION BY LASER BEAMS WITH DC AND HARMONIC INTENSITY COMPONENTS**

In the experimental configuration, the laser intensity included a dc component and even harmonics of 42 kHz, which is the driving frequency of the PEM. We first show the time evolutions of the $P_{2\Omega}$ and $\Delta T'$, a difference between temperatures at the top and bottom of the Co layer, as shown in Fig. 10(a) for $t_{Co}$ = 60 nm at $P_{2\Omega}$ = 52.1 mW. Furthermore, when the amplitude ratio of the dc, $2\Omega$, and $4\Omega$ of the $P$ waveform [Fig 10(b)] is set at 1:0.744:0.232 as in the experiments, we simulated the time evolution of $\Delta T'$. Figure 10(b) shows the time evolutions of $P$ and $\Delta T'$ for $t_{Co}$ = 60 nm at $P$ = 70 mW. In this case, the background temperature rises by approximately 10 K. Figure 10(c) indicates the fast Fourier transformation (FFT) of the $\Delta T'$ waveform shown in Fig. 10(b). The ratio of the FFT amplitudes for the dc, $2\Omega$, and $4\Omega$ components is 1:0.741:0.230, which is identical to the ratio of the laser amplitudes. Therefore, in the frequency range, even when the laser intensity includes the multiple frequency components, we can ignore the appearance of other frequency components in the $\Delta T'$ waveform.

**APPENDIX D: DERIVATION OF GEOMETRIC CORRECTION FACTOR**

Figure 11(a) shows a schematic of a Hall device illuminated by focused laser beams. For simplicity, we first consider that the spatial intensity profile of laser beams is top hat with a side length of $2\delta$. Figure 11(b) shows our model and the equivalent electrical circuit. Along the $z$-direction, a cuboid including the excited area is divided into $n$ pieces of strip with a size of $2\delta \times l \times \Delta t$. Here, the fractional resistance, $\Delta i$, of each small strip is expressed as follows:

$$\Delta i = \frac{Fj}{2\delta \Delta Z} \tag{7}$$

The combined resistance, $I^k$, of the other strips is given by

$$I^k = \frac{\Delta i}{l-1} \cong \frac{\Delta i}{l} = \frac{Fj}{2\delta Z} \tag{8}$$

$$I^{kk} = \frac{Fj}{(\ - 2\delta)Z} \tag{9}$$

$$I = \Delta i + \frac{\prime\ \prime\ \prime}{\prime + \prime\ \prime}. \tag{10}$$

$$\Delta!^{kk} = \left(\frac{\Delta[_{ANE}}{I}\right) \wedge \frac{I^{\ k}}{\prime + \prime\ \prime} a = \frac{I^{\ k}\Delta[_{ANE}}{\Delta i\ I^{\ k} + I^{\ k\prime}\ \prime + \Delta i\ \prime\ \prime}. \tag{11}$$

$$!^{kk} = \sum_{o\,p\,q}^{n} \Delta!_o^{kk} \tag{12}$$

$$= \frac{2 S_\text{ANE} \mu_0 H \sum_{k=1}^{n} \nabla T_k}{\rho_j L \left( \frac{1}{2 S \Delta Z} + \frac{1}{(L-2S) \rho Z} + \frac{1}{(L-2S) \Delta Z} \right)}$$

$$\cong \left( \frac{L-2S}{L} \right) \frac{2 w_0}{\rho_j} S_\text{ANE} \mu_0 H \sum_{k=1}^{n} \nabla T_k \Delta Z,$$

where $\nabla T_k$ is the vertical temperature gradient in a $k^\text{th}$ strip. Therefore, the measured electric voltage,

$$V_\text{ANE}' = E_y' w = \frac{2 w_0}{\rho Z} S_\text{ANE} \mu_0 H \sum_{k=1}^{n} \nabla T_k \Delta Z \tag{13}$$

Although the above derivation assumes that the intensity profile is top hat, we employed laser beams with a Gaussian spatial intensity profile in the experiments. Although the amplitude of the vertical temperature gradients depends on the position of the illuminated region, the laser-induced electric voltage depends linearly on the laser power, and the in-plane temperature gradients are canceled out. Therefore, we can calculate the electric voltages for excitation with a Gaussian laser beam using the following simple transformation. We consider the condition in which the total powers of the Gaussian and top-hat beams are identical, as follows:

$$\int_{-t}^{t} \int_{-t}^{t} I e^{-2(x^2+y^2)/r^2} \, dx \, dy = \int_{-x}^{x} \int_{-x}^{x} I' \, dx \, dy, \tag{14}$$

where $I$ and $I'$ denote the peak intensity of the Gaussian beam and intensity of the top-hat beam, respectively. In this case, the intensity ratio becomes $I : I' = 2 : 1$. Considering the intensity ratio, the geometrically corrected electric voltage, $V_\text{ANE}'$, becomes

$$V_\text{ANE}' = \frac{w_0}{2 \rho Z} S_\text{ANE} \mu_0 H \nabla T(A), \tag{15}$$

**APPENDIX E: IMPACT OF INTERFACIAL THERMAL RESISTANCE ON SIMULATIONS**

We performed simulations incorporating interfacial thermal resistances (ITR) at $t_\text{Co} = 60$ nm in the axisymmetric geometry. We used ITR values of $3.3 \times 10^{-6}$ K m$^2$ W$^{-1}$ at the Co/sapphire interface and $1.7 \times 10^{-7}$ K m$^2$ W$^{-1}$ at the Ru/Co interface, which are typical for metal/insulator and metal/metal interfaces, respectively [48]. Figure 12 shows the line profile of the temperature difference from the bottom of the Co layer along the thickness direction for $t_\text{Co} = 60$ nm. Although temperature jumps appear at each interface, the value of $\Theta/t_\text{Co}$ is scarcely affected ($\Theta/t_\text{Co} = 972$ K mm$^{-1}$). A similar trend was obtained for different $t_\text{Co}$ values. When temperatures at both ends are fixed and internal heat sources are absent, temperature gradients are strongly affected by the existence of ITR. In the present case, the presence of internal heat sources within the Co films likely account for the insensitivity of $\Theta/t_\text{Co}$ to ITR.

**APPENDIX F: HEATER EXPERIMENT AND CALCULATION OF ANOMALOUS NERNST EFFECT COEFFICIENT**

We demonstrate the characterization of the ANE using homogeneous temperature gradients prepared using a heater. The measurement configuration is shown in Fig. 13(a). We cut the Ru(5 nm)/Co (7 nm) film into a piece of 6 mm×3 mm, which was sandwiched between a Peltier heater and a 3-mm-thick AlN block of 5 mm×30 mm. The AlN block was fixed to a heat bath composed of oxygen-free Cu using Mo screws. When the temperature on the top surface of the Peltier heater was set at 60.0°C, the temperature on the top surface of the AlN block was 44.5°C. The temperatures were measured using thermistors. Under these temperature settings, we measured the ANE voltages as a function of the in-plane magnetic field [Fig. 13(b)]. The magnetic hysteresis shows a voltage jump of approximately 13 μV. Assuming that the temperature gradient was linear, we calculated the value of $\alpha_{ANE}$ to be approximately 0.22 μV K$^{-1}$ T$^{-1}$ using a temperature gradient of 4.6 K mm$^{-1}$.

Colossal anomalous Nernst effect in a correlated noncentrosymmetric kagome ferromagnet, Sci. Adv. **7**, eabf1467 (2021).

[16] M. Papaj, and L. Fu, Enhanced anomalous Nernst effect in disordered Dirac and Weyl materials, Phys. Rev. B **103**, 075424 (2021).

[17] I. Samathrakis, T. Long, Z. Zhang, H. K. Singh, and H. Zhang, Enhanced anomalous Nernst effects in ferromagnetic materials driven by Weyl nodes, J. Phys. D: Appl. Phys. **55**, 074003 (2022).

[18] T. Yamazaki, T. Seki, R. Modak, K. Nakagawara, T. Hirai, K. Ito, K. Uchida, and K. Takanashi, Thickness dependence of anomalous Hall and Nernst effects in Ni-Fe thin films, Phys. Rev. B **105**, 214416 (2022).

[19] H. Narita, M. Ikhlas, M. Kimata, A. A. Nugroho, S. Nakatsuji, and Y. Otani, Anomalous Nernst effect in a microfabricated thermoelectric element made of chiral antiferromagnet $Mn_3Sn$, Appl. Phys. Lett. **111**, 202404 (2017).

[20] T. Kikkawa, K. Uchida, S. Daimon, Y. Shiomi, H. Adachi, Z. Qiu, D. Hou, X.-F. Jin, S. Maekawa, and E. Saitoh, Separation of longitudinal spin Seebeck effect from anomalous Nernst effect: Determination of origin of transverse thermoelectric voltage in metal/insulator junctions, Phys. Rev. B **88**, 214403 (2013).

[21] T. C. Chuang, P. L. Su, P. H. Wu, and S. Y. Huang, Enhancement of the anomalous Nernst effect in ferromagnetic thin films, Phys. Rev. B **96**, 174406 (2017).

[22] K.-D. Lee, D.-J. Kim, H. Y. Lee, S.-H. Kim, J.-H. Lee, K.-M. Lee, J.-R. Jeong, K.-S. Lee, H.-S. Song, J.-W. Sohn et al., Thermoelectric Signal Enhancement by Reconciling the Spin Seebeck and Anomalous Nernst Effects in Ferromagnet/Non-magnet Multilayers, Sci. Rep. **5**, 10249 (2015).

[23] U. Martens, T. Huebner, H. Ulrichs, O. Reimer, T. Kuschel, R. R. Tamming, C.-L. Chang, R. I. Tobey, A. Thomas, M. Münzenberg et al., Anomalous Nernst effect and three-dimensional temperature gradients in magnetic tunnel junctions, Commun. Phys. **1**, 65 (2018).

[24] Y.-J. Chen and S.-Y. Huang, Light-induced thermal spin current, Phys. Rev. B **99**, 094426 (2019).

[25] C. Zhang, J. M. Bartell, J. C. Karsch, I. Gray, and G. D. Fuchs, Nanoscale Magnetization and Current Imaging Using Time-Resolved Scanning-Probe Magnetothermal Microscopy, Nano Lett. **21**, 4966 (2021).

[26] J. M. Bartell, D. H. Ngai, Z. Leng, and G. D. Fuchs, Towards a table-top microscope for nanoscale magnetic imaging using picosecond thermal gradients, Nat. Commun. **6**, 8460 (2015).

[27] A. Pandey, J. Deka, J. Yoon, A. Mathew, C. Koerner, R. Dreyer, J. M. Taylor, S. S. P. Parkin, and G. Woltersdorf, Anomalous Nernst Effect-Based Near-Field Imaging of Magnetic Nanostructures, ACS Nano **18**, 31949 (2024).

[28] M. Weiler, M. Althammer, F. D. Czeschka, H. Huebl, M. S. Wagner, M. Opel, I.-M. Imort, G.

Reiss, A. Thomas, R. Gross et al., Local Charge and Spin Currents in Magnetothermal Landscapes, Phys. Rev. Lett. **108**, 106602 (2012).

[29] R. Iguchi, S. Kasai, K. Koshikawa, N. Chinone, S. Suzuki, and K. Uchida, Thermoelectric microscopy of magnetic skyrmions, Sci. Rep. **9**, 18443 (2019).

[30] F. Johnson, J. Kimák, J. Zemen, Z. Šobáň, E. Schmoranzerová, J. Godinho, P. Němec, S. Beckert, H. Reichlová, D. Boldrin et al., Identifying the octupole antiferromagnetic domain orientation in $Mn_3NiN$ by scanning anomalous Nernst effect microscopy, Appl. Phys. Lett. **120**, 232402 (2022).

[31] H. Reichlova, T. Janda, J. Godinho, A. Markou, D. Kriegner, R. Schlitz, J. Zelezny, Z. Soban, M. Bejarano, H. Schultheiss et al., Imaging and writing magnetic domains in the non-collinear antiferromagnet $Mn_3Sn$, Nat. Commun. **10**, 5459 (2019).

[32] A. Yariv and P. Yeh, *Photonics: Optical Electronics in Modern Communications* (Oxford University Press, Oxford, UK, 2007).

[33] C. Multiphysics, Introduction to COMSOL Multiphysics Version: 6.2, *COMSOL Multiphysics* (Burlington, MA, 1998), accessed 17 December 2024.

[34] C. Multiphysics, Introduction to the Wave Optics Module Version: 6.2, *COMSOL Multiphysics* (Burlington, MA, 1998), accessed 17 December 2024.

[35] C. Multiphysics, Introduction to the Heat Transfer Module Version: 6.2, *COMSOL Multiphysics* (Burlington, MA, 1998), accessed 17 December 2024.

[36] J. M. D. Coey, *Materials for Spin Electronics*, in Spin Electronics, edited by M. Ziese and M. J. Thornton (Springer, Berlin, 2001).

[37] L. Ouarbya, A. J. Tosser, and C. R. Tellier, Effects of electron scatterings on thermal conductivity of thin metal films, J. Mater. Sci. **16**, 2287 (1981).

[38] A. D. Avery, S. J. Mason, D. Bassett, D. Wesenberg, and B. L. Zink, Thermal and electrical conductivity of approximately 100-nm permalloy, Ni, Co, Al, and Cu films and examination of the Wiedemann-Franz Law, Phys. Rev. B **92**, 214410 (2015).

[39] M. Ceccardi, A. Zeugner, L. C. Folkers, C. Hess, B. Büchner, D. Marré, A. Isaeva, and F. Caglieris, Anomalous Nernst effect in the topological and magnetic material $MnBi_4Te_7$, npj Quantum Mater. **8**, 76 (2023).

[40] K. Tang, Y. Yang, J. Shen, M. Shi, N. Zhang, H. Li, H. Li, Z. Liu, D. Shen, R. Wang et al., Unconventional anomalous Hall effect and large anomalous Nernst effect in antiferromagnet $SmMnBi_2$, Commun. Mater. **5**, 89 (2024).

[41] J. Xu, W. A. Phelan, and C.-L. Chien, Large Anomalous Nernst Effect in a van der Waals Ferromagnet $Fe_3GeTe_2$, Nano Lett. **19**, 8250 (2019).

[42] H. Yuasa, F. Nakata, R. Nakamura, and Y. Kurokawa, Spin Seebeck coefficient enhancement by using $Ta_{50}W_{50}$ alloy and YIG/Ru interface, J. Phys. Appl. Phys. **51**, 134002 (2018).

[43] G.-H. Park, H. Reichlova, R. Schlitz, M. Lammel, A. Markou, P. Swekis, P. Ritzinger, D. Kriegner, J. Noky, J. Gayles et al., Thickness dependence of the anomalous Nernst effect and the Mott relation of Weyl semimetal $Co_2MnGa$ thin films, Phys. Rev. B **101**, 060406 (2020).

[44] Y. Pu, D. Chiba, F. Matsukura, H. Ohno, and J. Shi, Mott Relation for Anomalous Hall and Nernst Effects in $Ga_{1-x}Mn_xAs$ Ferromagnetic Semiconductors, Phys. Rev. Lett. **101**, 117208 (2008).

[45] J. Weischenberg, F. Freimuth, S. Blügel, and Y. Mokrousov, Scattering-independent anomalous Nernst effect in ferromagnets, Phys. Rev. B **87**, 060406 (2013).

[46] D. Yang, L. Yi, S. Fan, X. He, Y. Xu, M. Liu, L. Ding, L. Pan, and J. Q. Xiao, Spin Seebeck coefficients of Fe, Co, Ni, and $Ni_{80}Fe_{20}$ 3d-metallic thin films, Mater. Res. Bull. **136**, 111153 (2021).

[47] J. D. Irwin and R. M. Nelms, *Basic Engineering Circuit Analysis*, 12th ed. (Wiley Publishing, New Jersey, 2020

[48] Y.-J. Wu, T. Zhan, Z. Hou, L. Fang, and Y. Xu, Physical and chemical descriptors for predicting interfacial thermal resistance, Sci. Data **7**, 36 (2020).

[49] J. R. Rumble, *CRC Handbook of Chemistry and Physics*, 104th ed. (CRC Press, Boca Raton, 2023).

[50] I. D. Marinescu, T. Doi, and E. Uhlmann, *Handbook of Ceramics Grinding and Polishing*, 2nd ed. (William Andrew, Norwich, 2015).

[51] D. A. Ditmars, S. Ishihara, S. S. Chang, G. Bernstein, and E. D. West, Enthalpy and Heat-Capacity Standard Reference Material: Synthetic Sapphire ($\alpha$-$Al_2O_3$) from 10 to 2250 K, J. Res. J. Res. Natl. Bur. Stand. **87**, 159-163 (1982).

[52] G. M. Katyba, K. I. Zaytsev, I. N. Dolganova, I. A. Shikunova, N. V. Chernomyrdin, S. O. Yurchenko, G. A. Komandin, I. V. Reshetov, V. V. Nesvizhevsky, and V. N. Kurlov, Sapphire shaped crystals for waveguiding, sensing and exposure applications, Prog. Cryst. Growth Charact. Mater. **64**, 133 (2018).

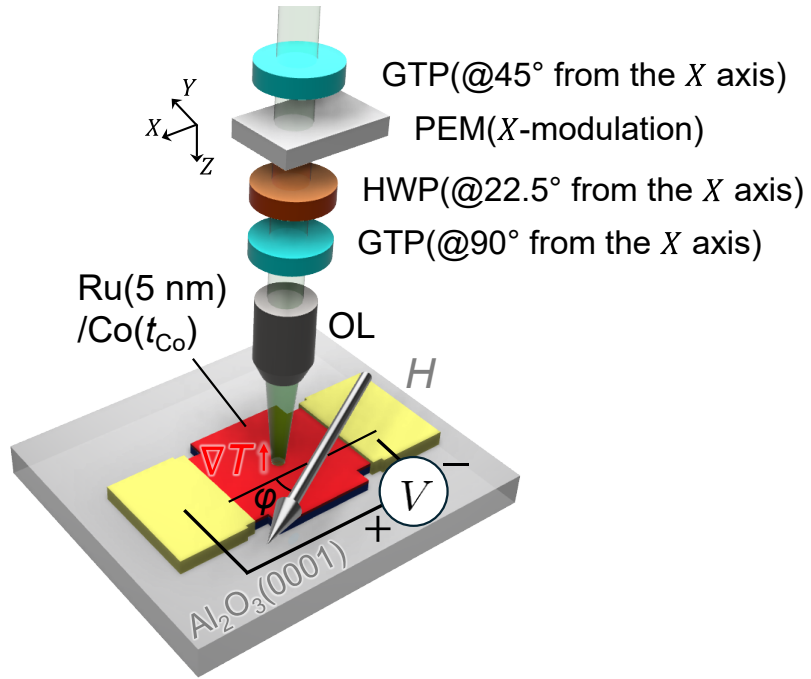

FIG. 1. Schematic of the experimental set-up. The intensity-modulated laser beams are focused on the center of the Hall cross, forming temperature gradients, $\nabla T$, along the thickness direction. In-plane magnetic field, $H$, is applied at an angle of $\varphi$ from the $X$ axis. Abbreviations: GTP (Glan Taylor prism), PEM (Photo-elastic modulator), HWP (Half-wave plate), and OL (Objective lens).

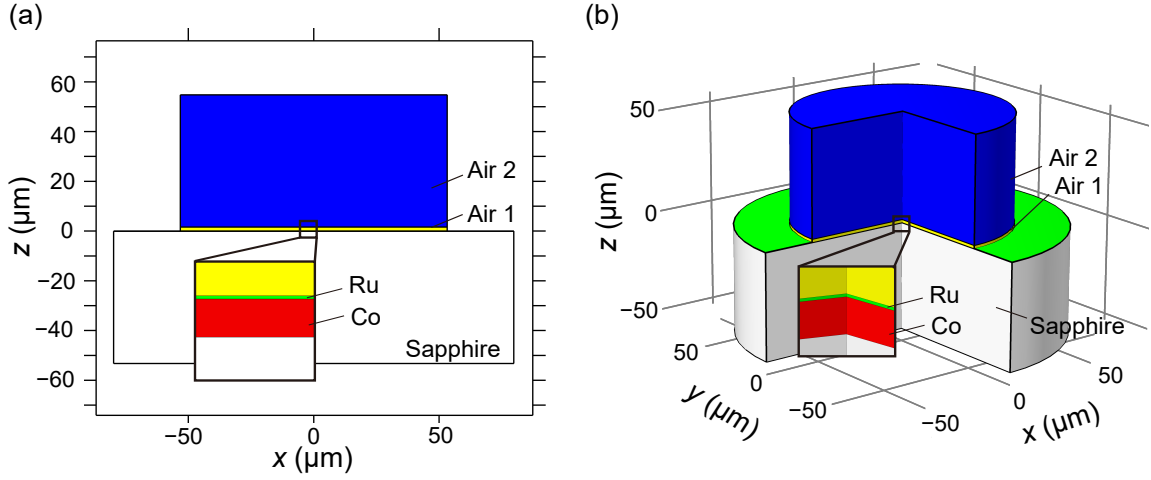

FIG. 2. (a) Two-dimensional model for finite-element simulation to calculate the spatial profile of the light electric field. We set the layer sizes of the substrate (white), Co (red), Ru (green), air 1 (yellow), and air 2 (blue) layers at $x \times z =$ 159.6 μm×53.2 μm, 159.6 μm×$t_{Co}$, 159.6 μm×5 nm, 106.4 μm×1.596 μm, and 106.4 μm×53.2 μm, respectively. For the simulation of the temperature profile, each layer was divided into a mesh. The $x$ and $z$ mesh sizes of the substrate, Co/Ru, air 1, and air 2 layers were 532 nm×532 nm, 532 nm×0.5 nm, 532 nm×31.92 nm, and 532 nm×532 nm, respectively. (b) Axisymmetric model to calculate the time evolution of temperature profile. The layer sizes and meshing method are identical to those used in the two-dimensional model.

TABLE 1. Material parameters used in the finite-element simulations. The densities, and heat capacities of Ru and Co and refractive indexes at $\lambda$ = 532 nm of Ru, Co, and sapphire are cited from ref. [49]. The thermal conductivities of Ru and Co are calculated from the measured resistivity based on Wiedemann-Franz's law. The density, heat capacity, and thermal conductivity of sapphire are cited from refs. [50–52], respectively.

| Material | Refractive index | Density [g cm$^{-3}$] | Heat Capacity [J kg$^{-1}$ K$^{-1}$] | Thermal Conductivity [W m$^{-1}$ K$^{-1}$] |
|---|---|---|---|---|
| Ru | $3.1 - 4.7i$ | 12.4 | 238 | $14.3 \pm 1.4$ |
| Co | $2.0 - 3.6i$ | 8.9 | 431 | $18.1 \pm 1.4$ |
| Sapphire | 1.77 | 4.0 | 750 | 41.9 |

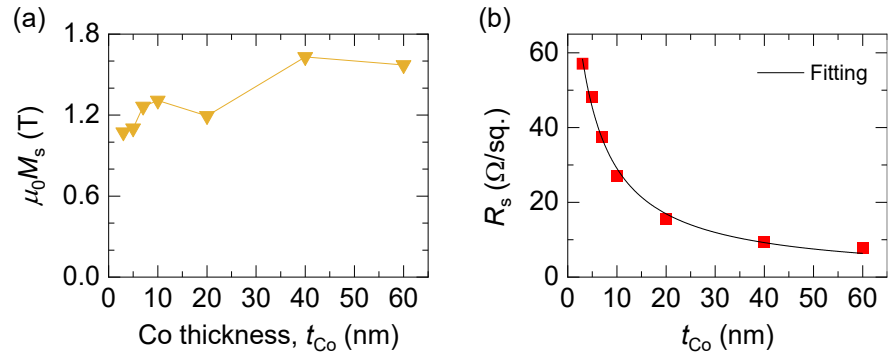

FIG. 3. Co thickness, $t_{Co}$, dependence of (a) saturation magnetization, $\mu_0 M_s$, and (b) sheet resistance, $R_s$. The fitting result with Eq. (5) is indicated by a solid black line.

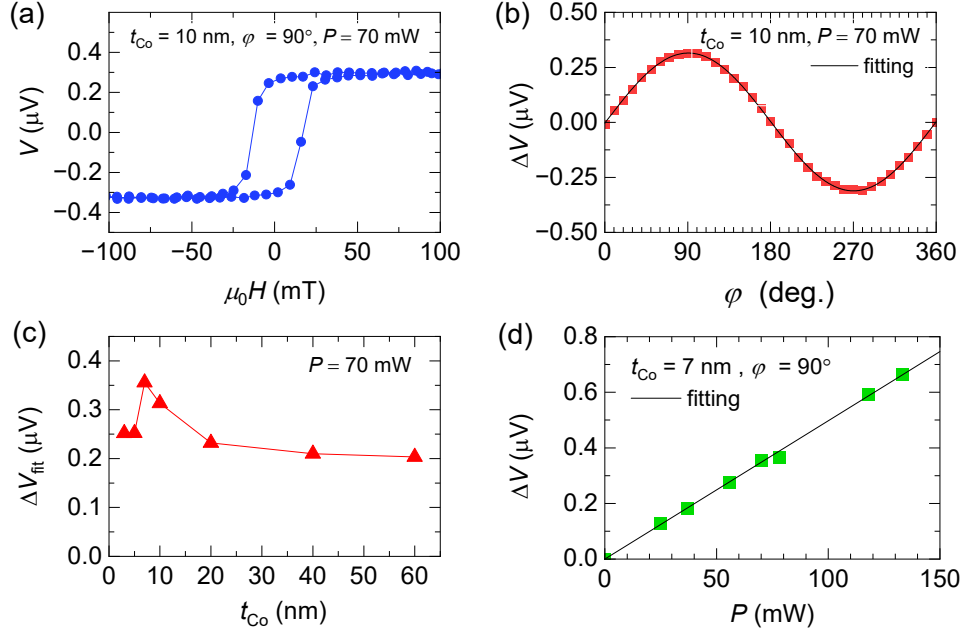

FIG. 4. Anomalous Nernst effect (ANE) induced by laser irradiation. (a) Magnetic hysteresis of electric voltages, $V$, measured for $t_{Co} = 10$ nm at $\varphi = 90°$ and an average laser power, $P$, of 70 mW. (b) $\varphi$ dependence of voltage jump, $\Delta V$, for $t_{Co} = 10$ nm at $P = 70$ mW. The $\varphi$ dependence at each $t_{Co}$ is fitted with a sine function $\Delta V_{fit}\sin\varphi$. (c) $t_{Co}$ dependence of $\Delta V_{fit}$ at $P = 70$ mW. (d) $P$ dependence of $\Delta V$ for $t_{Co} = 7$ nm at $\varphi = 90°$. We add a linear-fitting line to Fig. 4(d).

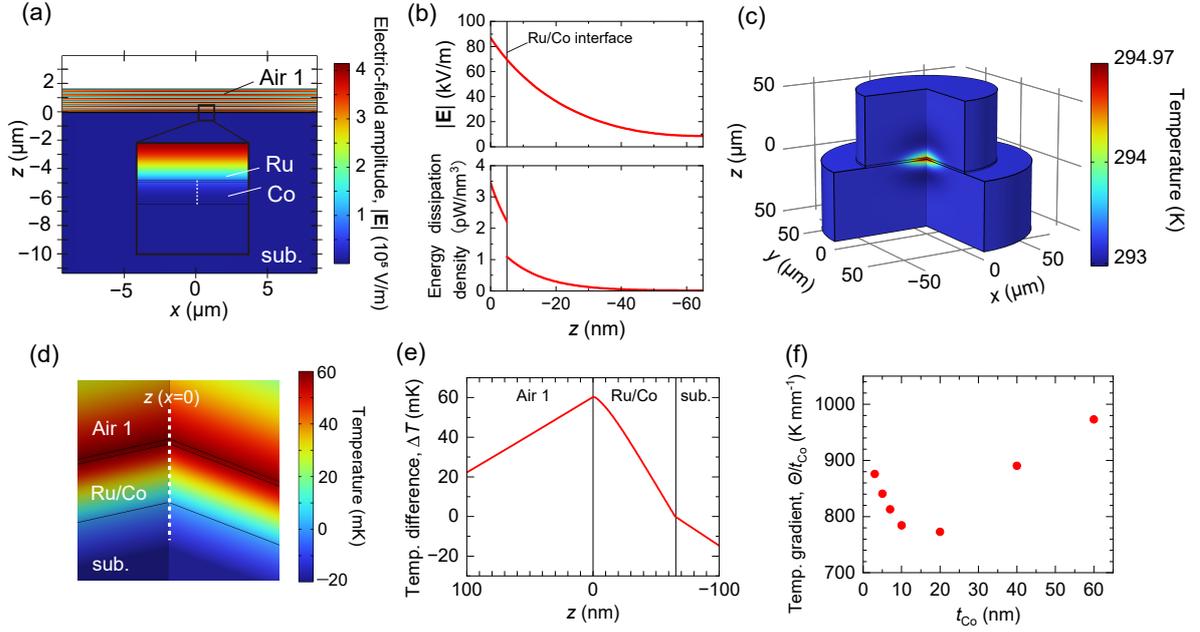

FIG. 5. Finite-element simulations of light electric field and temperature. (a) Spatial profiles of light electric-field amplitude, |E|, for $t_{Co}$ = 60 nm. A cut line is indicated by a white broken line. (b) Electric-field amplitude (top) and energy dissipation density (bottom) in the Ru/Co layer along the cut line. (c) Spatial profile and (d) magnified view of temperature for $t_{Co}$ = 60 nm when the temperature is maximized. The color scale in (d) shows the temperature variation from the bottom of the Co. A cut line is indicated by a white broken line. (e) Temperature difference, $\Delta T$, from the air1/Ru interface along the cut line. (f) $t_{Co}$ dependence of the integral of temperature gradients along the Co thickness, $\Theta$, normalized by $t_{Co}$. Throughout the finite-element simulations, we set the average power of the $2\Omega$ component at $P_{2\Omega}$ = 52.1 mW.

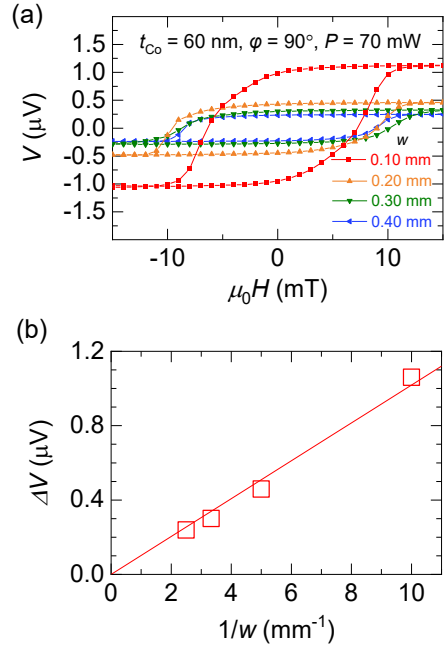

FIG. 6. Device-size dependence of ANE voltages. (a) Magnetic hysteresis of $V$ measured for $t_{Co} = 60$ nm at $\varphi = 90°$ and $P = 70$ mW. (b) Device-width dependence of $\Delta V$ for $t_{Co} = 60$ nm at $\varphi = 90°$. The solid line indicates the linear fitting.

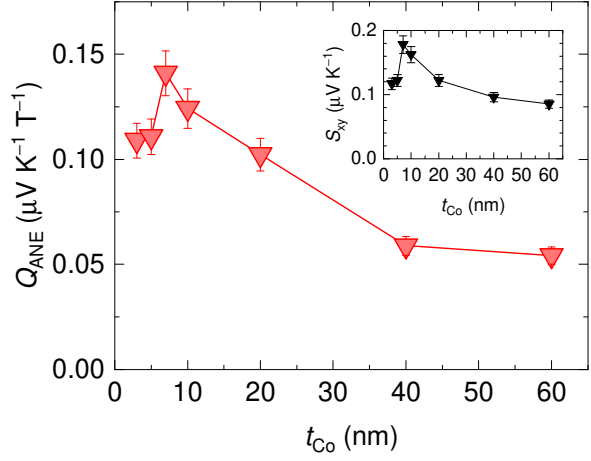

FIG. 7. $t_{Co}$ dependence of ANE coefficient, $Q_{ANE}$. The inset indicates the transverse Seebeck coefficient, $S_{xy}$, as a function of $t_{Co}$. The error bars were determined mainly by the uncertainty in the thermal conductivity of Ru and Co.

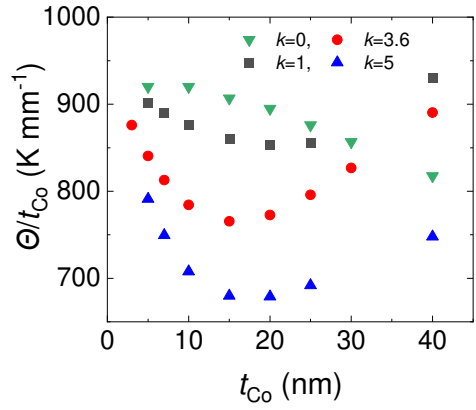

FIG. 8. $t_{Co}$ dependences of $\Theta/t_{Co}$ calculated with various extinction coefficients (imaginary part of the refractive indexes) of Co.

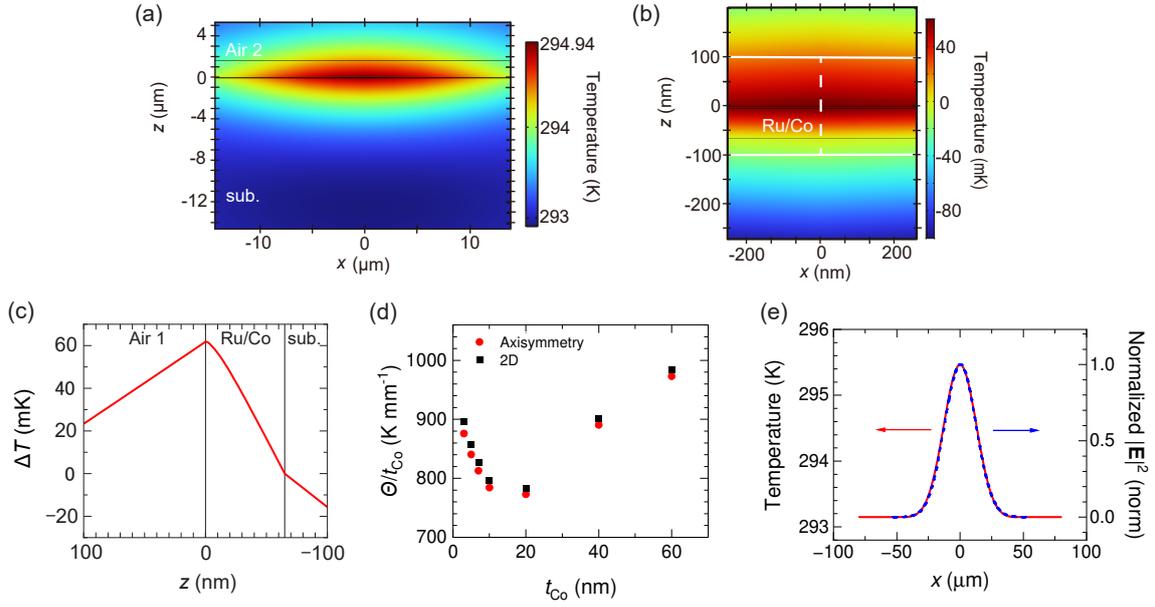

FIG. 9 (a) Spatial profile and (b) magnified view of temperature for $t_{Co}$ = 60 nm when the temperature is maximized calculated in the two-dimensional system. (c) Temperature difference, $\Delta T$, from the air1/Ru interface along the cut line. (d) Comparison of the $t_{Co}$ dependences of $\Theta/t_{Co}$ calculated using the two-dimensional systems with that calculated using axisymmetric geometry. (e) In-plane profiles of temperature (red) at $z$ = 35 nm and normalized squared electric-field intensity (blue) at the sample surface for $t_{Co}$ = 60 nm. Here, the in-plane temperature profile is plotted at the moment when the central temperature at $z$ = 35 nm reaches its maximum.

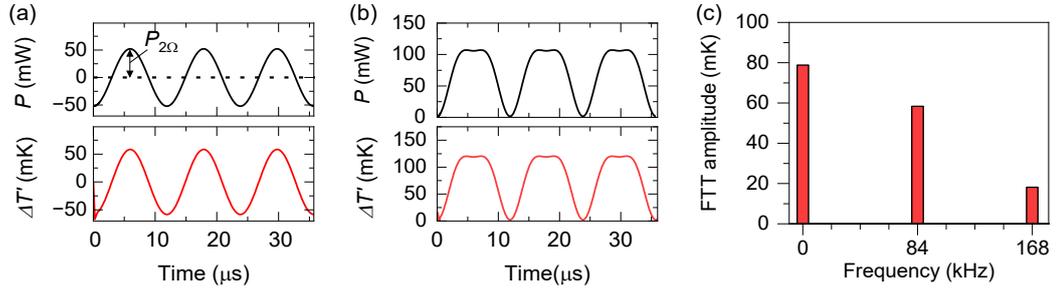

FIG. 10. Time evolution of temperature difference, $\Delta T'$, for $t_{Co}$ = 60 nm when the laser beam includes (a) $2\Omega$ amplitude component and (b) dc, $2\Omega$, and $4\Omega$ amplitude components. Here, the $\Delta T'$ is a difference between temperatures of the surface and bottom surface of the Co layer. (c) Amplitude of fast Fourier transformation (FFT) of the $\Delta T'$ waveform.

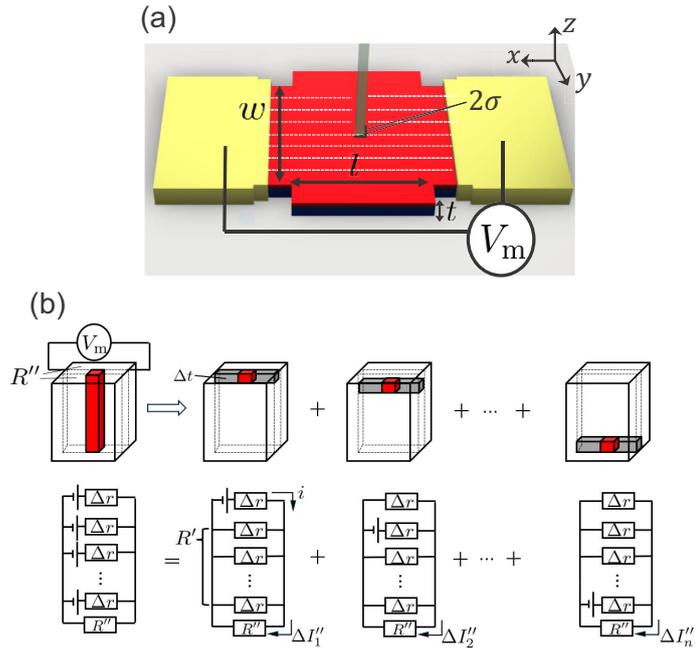

FIG. 11. (a) Schematic illustration for calculating geometrical correction factor. A laser beam, of which the spatial intensity profile is square with a side length of $2\sigma$ is focused on the center of a Hall cross with a size of $w \times l$. (b) Equivalent circuit model of ANE induced by local laser heating.

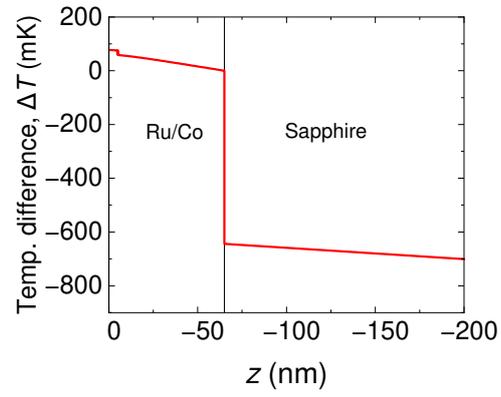

FIG. 12. Line profile of maximum temperature difference, $\Delta T$, at the spot center for $t_{Co} = 60$ nm when incorporating the interfacial thermal resistances at Ru/Co and Co/Sapphire interfaces. The other simulation conditions were identical to those for Fig. 5(c).

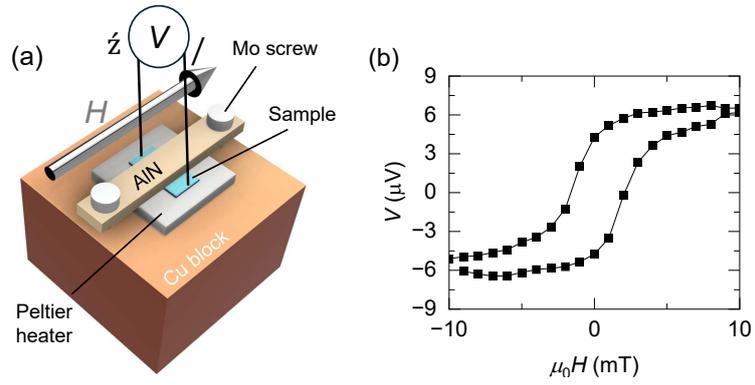

FIG. 13. Characterization of ANE using homogeneous temperature gradients prepared by a heater. (a) Schematic illustration of experimental configuration. (b) Magnetic hysteresis of electric voltage, $V$, for $t_{Co}$ = 7 nm. Here, we set the surface temperature of the Peltier heat at 60°C, when the surface temperature of the AlN block was 44.5°C.